\documentclass[journal]{IEEEtran}
\usepackage{mathrsfs}
\usepackage{pifont}
\usepackage{bbding}
\usepackage{amsmath,epsfig}
\usepackage{stmaryrd}
\usepackage{amssymb}
\usepackage{amsfonts}
\usepackage{epic}
\usepackage{graphicx}
\usepackage{subfigure}
\usepackage{curves}

\usepackage{algorithm}
\usepackage{algpseudocode}

\usepackage{stfloats}
\usepackage{latexsym}
\usepackage{epstopdf}
\usepackage{epic}
\usepackage{multirow}
\usepackage{stfloats}
\usepackage{bm}

\usepackage{booktabs}

\begin{document}

\newcommand{\reffig}[1]{Fig. \ref{#1}}

\newtheorem{definition}{Definition}
\newtheorem{theorem}{Theorem}
\newtheorem{corollary}{Corollary}
\newtheorem{lemma}{Lemma}

\renewcommand{\algorithmicrequire}{\textbf{Input:}}  
\renewcommand{\algorithmicensure}{\textbf{Output:}}  
\newenvironment{sequation}{\begin{equation}}{\end{equation}}
\newcounter{TempEqCnt}

\title{Position-Based Compressed Channel Estimation and Pilot Design for High-Mobility OFDM Systems}

\author{Xiang Ren, Wen Chen, \emph{Senior Member, IEEE},\\and Meixia Tao, \emph{Senior Member, IEEE}

\thanks{
Copyright (c) 2013 IEEE. Personal use of this material is permitted. However, permission to use this material for any other purposes must be obtained from the IEEE by sending a request to pubs-permissions@ieee.org.

The authors are with Department of Electronic Engineering,
Shanghai Jiao Tong University, Shanghai, China, 200240, (e-mail: \{renx, wenchen, mxtao\}@sjtu.edu.cn).
W. Chen is also with the school of Electronic Engineering and Automation, Guilin University of Electronic Technology.

This work is supported by the National 973 Project \#2012CB316106,
by NSF China \#61161130529, \#61328101, and \#61322102, by the STCSM Science and
Technology Innovation Program \#13510711200, and by the SEU National Key
Lab on Mobile Communications \#2013D11.
}}

\maketitle

\begin{abstract}

 With the development of  high speed trains (HST) in many countries, providing broadband wireless services in HSTs is becoming crucial. Orthogonal frequency-division multiplexing (OFDM) has been widely adopted for broadband wireless communications due to its high spectral efficiency.
 However, OFDM is sensitive to the time selectivity caused by high-mobility channels, which costs large spectrum or time resources to obtain the accurate channel state information (CSI).
Therefore, the channel estimation in high-mobility OFDM systems has been a long-standing challenge.
In this paper, we first propose a new position-based high-mobility channel model,
in which the HST's position information and Doppler shift are utilized to determine the positions of the dominant channel coefficients. 
Then, we propose a joint pilot placement and pilot symbol design algorithm for compressed channel estimation. It aims to reduce the coherence between the pilot signal and the proposed channel model, and hence can improve the channel estimation accuracy.
Simulation results demonstrate that the proposed method achieves better performances than existing channel estimation methods over high-mobility channels. Furthermore, we give an example of the designed pilot codebook to show the practical applicability of the proposed scheme.

\end{abstract}


\section{Introduction}
Orthogonal frequency-division multiplexing (OFDM) has been widely adopted for broadband wireless communication systems due to its high spectral efficiency \cite{1}. In OFDM systems, each subcarrier has a narrow bandwidth which ensures the signal robust against the frequency selectivity caused by the multipath delay spread. 
However, OFDM is sensitive to the time selectivity, which is induced by rapid time variations of mobile channels.
In recent years, high speed trains (HST)  have been increasingly developed in many countries and OFDM has been adopted for high data rate services \cite{5}. Since an HST travels at a speed of around 500 km/h, the wireless channels suffer from a high Doppler shift.
In high-mobility environments, wireless channels are both fast time-varying and frequency selective and can be considered as the doubly selective channels \cite{2}-\cite{4}.
As the quality of channel estimation has a major impact on the overall system performance, it is necessary to investigate  reliable estimation methods in high-mobility environments.

Channel estimation  for fast time-varying channels has been extensively studied in the literature, 
and various time-varying channel models have been established.
The works \cite{90} and \cite{91} proposed several channel estimators by using a linear time-varying channel model. They assumed that the channel varies with time linearly in one or more OFDM symbols. 
The method in \cite{90} works well at low Doppler shifts since some channel matrix coefficients are ignored.
The work \cite{91} proposed two approaches to estimate time-varying channels: one uses guard intervals and the other exploits three consecutive symbols.
However, the linear models can result in large modeling error and severely degrade the channel estimation performance in high-mobility environments, where the channel may change significantly even within one OFDM symbol.
To overcome the modeling problem, the authors in \cite{16}-\cite{42} proposed several basis expansion models (BEM) for the time variations of each OFDM symbol.
The work \cite{16} assumed a polynomial BEM channel model and suggested an iterative channel estimation method.
The work \cite{6} proposed channel estimators based on a windowed BEM to combat both the noise and the out-of-band interference.
In particular, the work \cite{42} claimed that the equidistant pulse-shaped pilot placement is optimal for a BEM-based doubly selective channel.
These works are based on the implicit assumption of  rich underlying multipath environments.

Recently, growing experimental studies have shown that the high-mobility channels  tend to exhibit a sparse structure at some high dimensional signal spaces, such as the delay-Doppler domain, and can be characterized by significantly fewer parameters.
To utilize the inherent channel sparsity, the authors in \cite{2}-\cite{4}, \cite{10} and \cite{11}  studied the applications of compressed sensing (CS) in doubly-selective channels, which well reflect the natures of the high-mobility channels.
The works \cite{2}-\cite{4} introduced the notion of channel sparsity and presented CS-based approaches to estimate the channel state information (CSI).
The works \cite{10} and \cite{11} optimized the delay-Doppler basis to  improve the estimation performances.

\begin{figure*}[!t]
\centering
\includegraphics[width=5.3in]{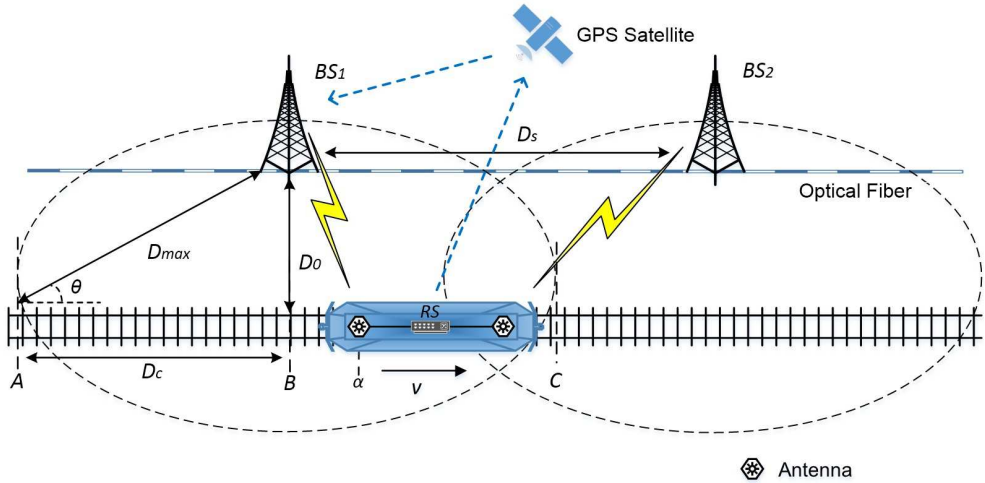}
\centering
\caption{The structure of the HST communication system. }\label{fig1}
\end{figure*}

Coherence is an important issue in CS and fundamental researches \cite{12}-\cite{15} show that the coherence influences the CS reconstruction performance directly.
The works \cite{12}-\cite{14} concluded that a lower coherence between the measurement matrix and the dictionary matrix in CS leads to a better performance.
Therefore, how to design the pilot signal to reduce the system coherence in a high-mobility environment is a very interesting and valuable problem.
Previous works \cite{20}-\cite{26} studied several pilot design methods to minimize the CS coherence and improve the system performance.
The works \cite{20}-\cite{22} proposed several pilot placement design methods based on a exhaustive search to reduce the system coherence in sparse channels with large iterations.
The works \cite{23} and \cite{24} designed the pilot placement  by using a discrete stochastic approximation method for sparse OFDM channels without considering the channel mobility and the inter-carrier interference (ICI).
In the previous work \cite{25}, we designed the pilot symbol by shrinking the Gram matrix  in a high-mobility multiple-input multiple-output (MIMO) OFDM system.
However, none of these works considered the joint optimization of the pilot symbol and pilot placement.

In this paper,  we introduce a new position-based compressed channel estimation method for OFDM systems over high-mobility channels, in which the pilot placement and pilot symbol are jointly designed to further improve the performance.
In specific, we propose a new position-based high-mobility channel model which reflects the Doppler shift according to the HST position. The HST's position information and Doppler shift are utilized to predict the positions of the sparse dominant channel coefficients, which highly reduces the estimation complexity.
Then, based on the CS coherence minimization criterion, a joint pilot placement and pilot symbol design algorithm is proposed to reduce the coherence between the pilot signal and the proposed channel model.
Simulation results demonstrate that the proposed method achieves better performances than existing channel estimation methods in the  high-mobility environment. Furthermore, we give an example of a designed pilot codebook to show the practical applicability of the proposed scheme.

The rest of this paper is organized as follows. Section II introduces the HST communication model, the OFDM system model, and the conventional high-mobility channel model. In Section III, a new position-based high-mobility channel model is proposed. Then we introduce a new position-based channel estimation method and discuss the ICI mitigation method. In Section IV,  we propose a joint pilot placement and pilot symbol design algorithm and discuss its complexity and convergence. The practical applicability of the proposed scheme is also discussed.  Section V presents simulation results in the high-mobility environment. Finally, Section VI concludes this paper.

$Notations$: 
$\left\|\cdot\right\| _{\ell _0 }$ denotes the number of nonzero entries in a matrix or vector, and $\left\|\cdot\right\| _{\ell _2 }$ is the Euclidean norm. Notation $\phi (k,u)$ denotes the $(k,u)$-th entry of the matrix ${\bf{\Phi}}$, and $\phi(:,u)$ denotes the $u$-th column vector of the matrix $\bf{\Phi}$. The superscripts $(\cdot)^T$ and $(\cdot)^H$ denote the transposition and Hermitian of a matrix, respectively. $\otimes$ denotes the Kronecker product, and ${\bf{a}} = vec\{\bf{A}\}$ denotes the vector obtained by stacking the columns of matrix $\bf{A}$.
Finally, $\mathbb{R}$ denotes the real field and  $\mathbb{Z}$ denotes the set of integers.


\section{System Model}
\subsection{High Speed Train Communications}

We consider a well-recognized system architecture of broadband wireless communications for  high speed trains (HST) \cite{5}\cite{9}, as shown in Fig.~1.  The communication between base stations (BS) and mobile users is conducted in a two-hop manner through a relay station (RS) deployed on the train.
The RS has two antennas on the top of the train to communicate with the BS. On the other hand, multiple indoor antennas are distributed in the train carriages to communicate with mobile users by existing wireless communication technologies, e.g. wireless fidelity (WiFi). The BSs are located 10 to 50 meters away from the railway at some intervals and connected with optical fibers. Here we assume each BS is equipped with one antenna and has the same power allocation and coverage range.

In this paper, we  focus on the channel estimation  between the BSs and the RS on the HST.
When the HST camps in a single cell, the RS selects the antenna with better channel quality to communicate with the BS; when the HST moves across the cell edges, the front antenna executes handover while the rear one keeps connect to the serving BS.
The HST is equipped with a global positioning system (GPS) which can estimate the HST's instant position and speed information and send them to the BS \cite{92}. Several factors may influence the performance of the GPS, such as signal arrival time measurement, atmospheric effects, terrains and so on. Particularly, when the HST runs in a tunnel, the GPS accuracy may be highly reduced.
However, in this paper, we do not consider these factors and assume that the HST runs in a plain terrain.
We also assume that the GPS estimates the HST's speed and position information perfectly and send to the BS with no time delay.

Denote $v$ as the speed of the HST and $c$ as the light speed. The distance between BSs is denoted as $D_s$. Let $D_{max}$ denote the maximum distance of the coverage of the BS to the railway, i.e. the position $A$ and $C$ to $BS_1$. Let $D_{0}$ denote the minimum distance, i.e. the position $B$ to $BS_1$, and $D_c$ denote the distance between $A$ and $B$. In each cell, we define the HST position $\alpha$ as the distance between the serving antenna and the position $A$, and $\alpha = 0 $ at $A$.
Let $\theta$ denote the angle between the signal transmitted from the BS to RS and the railway.
When the HST moves from $A$ to $C$,  $\theta$  changes  from $\theta_{min}$ to $\theta_{max}$.
If $D_{max} \gg D_{0}$, then we have $\theta_{min}\approx 0^\circ$ and $\theta_{max}\approx 180^\circ$.
Furthermore, HST suffers from the Doppler shift $f_d$  at different positions, and $f_d$  can be calculated by using the equation $f_d = \frac{v}{c}\cdot f_c \cos\theta$, in which $f_c$ is the carrier frequency.  It is easy to find that $f_{{d}_0} = 0$ at  $B$ for $\theta = 90^\circ$.
%

\subsection{OFDM System}

We consider an OFDM system with $K$ subcarriers for the link between the BS and the RS in the HST communication system. The transmit signal at the $k$-th subcarrier of the $n$-th OFDM symbol is denoted as $X^n (k)$, for $n = 1,2,...,N$ and $k = 1,2,...,K$. The transmitter performs the inverse discrete Fourier transform (IDFT), inserts the cyclic prefix (CP), and then transmits the signals to the channel.
After removing the CP and passing the DFT operation at the receiver, the received signal in the frequency domain can be represented as
\begin{equation}
\mathbf{Y}^n  =\mathbf{H}^n\mathbf{X}^n+\mathbf{W}^n\label{eq0},
\end{equation}
where ${\bf{Y}}^n = [Y^n(1),Y^n(2),...,Y^n(K)]^T$ is the received signal vector over all subcarriers during the $n$-th OFDM symbol, ${\bf{H}}^n$ is a $K\times K$ channel matrix in the frequency domain,
${\bf{X}}^n= [X^n(1),X^n(2),...,X^n(K)]^T$ is the transmitted signal vector over all subcarriers,
and ${\bf{W}}^n = [W^n(1),W^n(2),...,W^n(K)]^T$ denotes the noise vector, where $W^n(k)$ is the additive white Gaussian noise (AWGN) with a zero mean and $\sigma^2_W$ variance.
The entries of ${\bf{H}}^n$ are represented as
\begin{align}
{{H}^n({k,d}})&=\frac{1}{K}\sum\limits_{m=0}^{K-1}{\sum\limits_{\ell=0}^{I-1}{{{h}}^n\left(\ell,m \right){{e}^{-j\frac{2\pi }{K}\ell(k-1)}}}}{{e}^{j\frac{2\pi }{K}\left( d-k \right)m}}, \nonumber
\\&~1\leq k,d\leq K, \label{eq1}
\end{align}
where ${{h}}^n\left(\ell,m \right)$ is the $\ell$-th channel tap in the $m$-th sample time of the $n$-th OFDM symbol, and $I$ is the maximum number of channel taps.
More detailed descriptions of ${\bf H}^n$ in high-mobility environments is given in the next subsection.

If the channel is time-invariant, 
the off-diagonal term ${{H}^n({k,d}})$ $(k\neq d)$ is negligible, and the diagonal term ${{H}^n({k,d}})$ $(k=d)$ alone represents the channel in the frequency domain.
Therefore, the channel matrix ${\bf{H}}^n$ can be divided into two parts as the ICI-free channel matrix ${\bf{H}}_{free}^n \triangleq diag\{[H^n(1,1),H^n(2,2),...,H^n(K,K)]\}$ and the ICI channel matrix ${\bf{H}}_{ICI}^n \triangleq {\bf{H}}^n - {\bf{H}}_{free}^n$. Then (\ref{eq0}) can be rewritten as
\begin{align}
{\bf{Y}}^n  &={\bf{H}}_{free}^n{\bf{X}}^n + {\bf{H}}_{ICI}^n{\bf{X}}^n + {\bf{W}}^n,\\
&= {{\bf{X}}^n_{d}}{\bf{H}}_{vec}^n + {\bf{H}}_{ICI}^n{\bf{X}}^n + {\bf{W}}^n,\label{eq20}
\end{align}
where $ {{\bf{X}}^n_{d}} = diag\{ [X^n(1),X^n(2),...,X^n(K)]^T\}$ is a diagonal matrix of ${\bf{X}}^n$, and ${\bf{H}}_{vec}^n = vec\{{\bf{H}}^n_{free}\}$ is the stacking vector of ${\bf{H}}^n_{free}$.

\subsection{High-Mobility Channel Model}
Let $\tau _{\max }$ be the maximum delay spread, $f_{d_{\max }}$ be the maximum Doppler shift, $T_d$ be the packet duration and $W$ be the system bandwidth. Denote $T_0$ as the OFDM symbol duration and $W_0$ as the bandwidth of each subcarrier, $N_t = T_d/T_0$ and $N_f = W/W_0$. The high-mobility channel at the $k$-th subcarrier of the $n$-th OFDM symbol  in the delay-Doppler domain \cite{2}-\cite{4} can be modeled as
\begin{equation}
H(n,k) = \sum\limits_{l = 0}^{L - 1} {\sum\limits_{m =  - M}^M {\beta}_{l,m} e^{j2\pi \frac{m}{N_t}n} e^{ - j2\pi \frac{l}{N_f}k} },\label{eq00}
\end{equation}
where $L = \lceil{W\tau_{\max}}\rceil+1$ represents the maximum number of resolvable paths and $M = \lceil{2T_d  f_{d_{\max }}}\rceil$ represents the maximum number of resolvable Doppler shifts, 
$\beta_{l,m}$ is the  channel coefficient of the $l$-th resolvable path with the resolvable Doppler shift $m$.

For the sake of convenience, we define two vectors ${\bf{u}}_k  = \left[ 1 , {e^{ - j2\pi \frac{1}{N_f}k} } ,{...} , {e^{ - j2\pi \frac{(L-1)}{N_f}k} }  \\
 \right]$ and ${\bf{u}}_n = \left[
   {e^{j2\pi \frac{{ - M}}{N_t}n} } , {e^{j2\pi \frac{{( - M + 1)}}{N_t}n} } , {...} , {e^{j2\pi \frac{M}{N_t}n} }  \\
 \right]$.
 Then the channel model can be represented as a matrix form:
\begin{equation}
H(n,k)= {\bf{u}}_k {\bf{B}}{\bf{u}}^T_n  = \left( {{\bf{u}}_n  \otimes {\bf{u}}_k } \right) {\bf{b}}
\label{eq2},
\end{equation}
 where ${\bf{B}}$ is an $L \times (2M+1)$ channel coefficient matrix in the delay-Doppler domain, i.e.,
\begin{equation}
\mathbf{B}\triangleq\left[ \begin{matrix}
   {{\beta }_{0,-M}} & \cdots  & {{\beta }_{0,0}} & \cdots  & {{\beta }_{0,M}}  \\
   \vdots  & \ddots  & \vdots  & \ddots  & \vdots   \\
   {{\beta }_{L-1,-M}} & \cdots  & {{\beta }_{L-1,0}} & \cdots  & {{\beta }_{L-1,M}}  \\
\end{matrix} \right],
\end{equation}
and ${\bf{b}} \triangleq vec\{{\bf B}\}$ is the stacking vector of $\bf B$, i.e.,
\begin{align}
{\bf{b}}
&= \left[\begin{matrix}
{\bf{b}}^T_{-M},\ldots,{\bf{b}}^T_0,\ldots,{\bf{b}}^T_{M}\\
\end{matrix}\right]^T,\\
&= \left[\begin{matrix}
 { {\beta }_{0,-M}},\ldots,{{\beta }_{L-1,-M}},\ldots,{{\beta }_{0,M}},\ldots,{{\beta }_{L-1,M}}\\
\end{matrix}\right]^T,
\end{align}
where ${\bf{b}}_x$ denotes the column vector of ${\bf{B}}$ and $x = -M,-M+1,...,M$.

To explore the sparsity of the high-mobility channel, here we introduce the definition of $S$-sparse channel based on \cite{3}.
In general, the high-mobility channel is $S$-sparse in the delay-Doppler domain, due to the large number of the non-dominant channel coefficients \cite{4}.
In this paper, we assume that the coefficients are constant within each OFDM symbol and different for different symbols.
\begin{definition}[$S$-sparse Channels \cite{3}]\label{df1}
Define the dominant coefficients of a wireless channel as the coefficients which contribute significant powers, i.e. $|{\beta _{l,m}}|^2 > \gamma$, where $\gamma$ is a chosen threshold.
We say that the channel is  $S$-sparse if  the number of its dominant coefficients satisfies $S = \left\| {{\bf{b}}} \right\|_{\ell _0 }  \ll N_0 = L(2M+1)$, where $N_0$ is the total number of the channel coefficients.
\end{definition}

\section{Position-Based High-Mobility Channel Estimation}\label{se3}

In this section, we first propose a new position-based channel model to reduce the number of channel estimation parameters by utilizing the position information. Then, based on the proposed channel model, we give a position-based channel estimation scheme. Finally, the ICI mitigation scheme is also discussed.

\subsection{Position-Based High-Mobility Channel Model}


Considering the HST communication system model shown in Fig.~1. 
When the HST is at a certain position $\alpha$ with speed $v$, the high-mobility channel suffers from a certain Doppler
shift $f_d$.  Here we  assume that $v$ is constant during the HST passing a cell, and $f_d$ is constant in one OFDM symbol. In this case, it can be found that the dominant coefficients only exist in the ${\bf{b}}_x$ for suffering the same $f_d$, which can be represented as
 \begin{equation}
{\bf{b}}_x = 
\left[\begin{matrix}
   {{\beta_{0,x}}},{{\beta_{1,x}}},\ldots,{{\beta_{L-1,x}}}\\
\end{matrix}\right]^T.
\end{equation}
This is reasonable since ${\bf{b}}_x$ represents the $L$ resolvable paths with the resolvable Doppler shift $m$.
The relationship between the subscript $x$ and $f_d$ is given as
\begin{align}
x=\left\{\begin{matrix}
  &\left\lceil2{T_d}{f_d}\right\rceil,~&f_d\in\left[0,f_{d_{max}}\right],\\
  &\left\lfloor2{T_d}{f_d}\right\rfloor,~&f_d\in\left[-f_{d_{max}},0\right).
\end{matrix}\right.\label{eq12}
\end{align}
Denote $\tilde{M} =2T_d  f_{d_{\max }} =2T_d\frac{v}{c} \cdot f_c$. Then the relationship between $x$ and $\alpha$ can be represented as
\begin{align}
x=\left\{\begin{matrix}
  &\left\lceil\tilde{M}\cdot\frac{D_{c}-\alpha}{\sqrt{{(D_{c}-\alpha)}^2+{D_0}^2}}\right\rceil,~&{\alpha}\in[0,D_{c}],\\ \\
   &\left\lfloor\tilde{M}\cdot\frac{D_{c}-\alpha}{\sqrt{{(D_{c}-\alpha)}^2+{D_0}^2}}\right\rfloor,~&{\alpha}\in(D_{c},2D_{c}].
\end{matrix}\right.\label{eq13}
\end{align}
Note that  when $D_{max} \gg D_{0}$, we have $D_{c} = D_{max}$.

\begin{figure}[!t]
\centering
\includegraphics[width=3.5in]{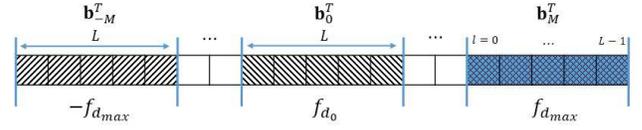}
\caption{The structure of the coefficient vector ${\bf{b}}^T$. }\label{fig2}
\end{figure}

The structure of the coefficient vector ${\bf{b}}^T$ is shown as \reffig{fig2}.
In particular,
${\bf{b}}^T_M$ $(x=M)$ with the cross lines denotes the column vector including the dominant coefficients with $f_{d_{max}}$ at the position $A$, ${\bf{b}}^T_0$ $(x=0)$ with the slash lines denotes the one with $f_{d_0} = 0$ at $B$, and ${\bf{b}}^T_{-M}$ $(x=-M)$ with the back slash lines denotes the one with $-f_{d_{max}}$ at  $C$, respectively.
When the HST moves from $A$ to $C$,  ${\bf{b}}^T_x$ changes from ${\bf{b}}^T_M$ to ${\bf{b}}^T_{-M}$ in sequence. On the other hand, other coefficients are non-dominant which contribute less to the CSI, according to Definition \ref{df1}.

\emph{Remark}: (\emph{Channel Sparsity}) It is easy to find that, with a certain $f_d$, ${\bf{b}}_x$ contains at most $L$ dominant coefficients and  the sparsity is $S$, i.e. $\left\| {{\bf{b}}}_x \right\|_{\ell _0 } =\left\| {{\bf{b}}} \right\|_{\ell _0 } = S\leq L < L(2M+1)$. Furthermore, high-mobility channels are considered as the doubly-selective channels in \cite{2}-\cite{4}  and have the multipath sparsity, which means there are only $S$ paths $(S\ll L)$ with large coefficients and others can be neglected.
In addition, as $M$ increases with the Doppler shift caused by the HST speed, high-mobility will introduce a large $M$. Therefore, we have $\left\| {{\bf{b}}}_x \right\|_{\ell _0 }  = S\ll L \ll L(2M+1)$, and the high-mobility channel is $S$-sparse in the proposed position-based model.

After knowing the position of ${\bf b}_x$ which includes the dominant coefficients, we can further get the dominant channel model ${\bf{\Phi}}_x$.
Let ${\bf{\Phi }} = \left[{ {\bf{u}}_n } \otimes {\bf{u}}_{k_1 }; {{\bf{u}}_{n }  \otimes {\bf{u}}_{k_2} };  \cdots ; {{\bf{u}}_{n }  \otimes {\bf{u}}_{k_P} }  \right]$ be the $P\times L(2M+1)$  channel model dictionary matrix of the $n$-th OFDM symbol with $P$ pilots, in which ${\bf{u}}_{k_p} = {\bf{u}}_{k|k = k_p}$ and the pilot placement is  ${\bf{p}} = [k_{1},k_{2},...,k_{P}]$.
Denote ${\bf{\Phi}}_x$ as the $P\times L$ dominant channel model whose columns are corresponding to  ${\bf{b}}_x$. Then the dominant channel model ${{\mathbf{\Phi }}_{x}}$ can be represented as
 \begin{equation}
 {{\mathbf{\Phi }}_{x}}=\left[ \begin{matrix}
   \phi \left({{k}_{1}},\omega_x\right)& \cdots & \phi \left({{k}_{1}},\omega_x+L-1\right)  \\
   \vdots  & \ddots  & \vdots   \\
   \phi ({{k}_{P}},\omega_x)& \cdots & \phi ({{k}_{P}},\omega_x+L-1)  \\
\end{matrix} \right],
\end{equation}
where $\omega_x = L(M+x)$. Note that the columns of ${{\mathbf{\Phi }}_{x}}$ represents the  resolvable paths and the rows represents the pilot subcarriers.

\begin{figure}[!t]
\centering
\includegraphics[width=3.5in]{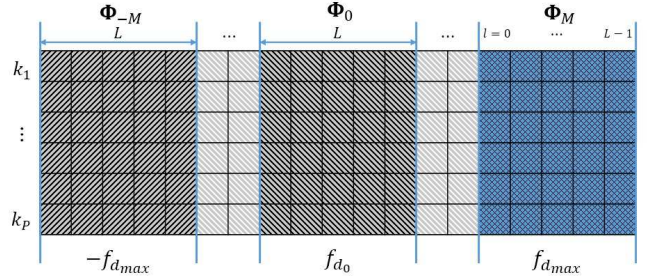}
\caption{The structure of the high-mobility channel model matrix ${\bf{\Phi}}$. }\label{fig3}
\end{figure}

Similarly, the structure of ${\bf{\Phi }} = [{{\bf{\Phi}}}_{-M},...,{{\bf{\Phi}}}_{0},...,{{\bf{\Phi}}}_{M}]$ is shown as \reffig{fig3}.
In particular, ${\bf{\Phi}}_M$ with the cross lines denotes the dominant channel model with $f_{d_{max}}$ at position $A$ corresponding to ${\bf b}_M$,
${\bf{\Phi}}_0$  with the slash lines denotes the one with $f_d=0$ at  $B$,
and ${\bf{\Phi}}_{-M}$  with the back slash lines denotes the one with $-f_{d_{max}}$ at $C$, respectively.
When the  HST moves from $A$ to $C$, ${\bf{\Phi}}_x$ changes from ${\bf{\Phi}}_M$ to ${\bf{\Phi}}_{-M}$ in sequence. More specifically, as the dominant coefficients only exist in ${\bf{b}}_x$ corresponding to a certain Doppler shift $f_d$ or HST position $\alpha$, we only need to consider the ability of  ${\bf{\Phi}}_x$ and estimate ${\bf{b}}_x$, which highly reduce the analysis and computational complexity.


\subsection{Position-Based Channel Estimation}
Using the proposed position-based channel model in the previous subsection, we propose a channel estimation method based on the comb-type pilot.
Assume that there are  $P$ ($P\leq K$) pilots and insert at the pilot placement set ${\bf{p}} = [k_{1},k_{2},...,k_{P}]$. The pilot placement and pilot symbols are fixed during one OFDM symbol.
Since we only consider the system in one OFDM symbol, the superscripts $n$ in the rest of the paper are omitted for compactness. Then, based on the proposed position-based channel model, Eq.~(\ref{eq20}) of the received pilot vector can be represented as follows:
\begin{align}
 {\bf{Y}}({\bf{p}}) &={\bf{X}}_{d}({\bf{p}}){\bf{H}}_{vec}({\bf{p}})+{\bf{d}}+{\bf{W}}({\bf{p}}), \\
 &={\bf{X}}_{d}({\bf{p}}){{\bf{\Phi}}}{\bf{b}}+{\bf{d}}+{\bf{W}}({\bf{p}}),\\
&={\bf{X}}_{d}({\bf{p}}){{\bf{\Phi}}_x}{\bf{b}}_x+{\bf{d}}+{\bf{W}}({\bf{p}}),\label{eq8}
\end{align}
in which ${\bf{Y}}({\bf{p}})=[Y(k_1),Y(k_2),...,Y(k_P)]^T$ denotes the received pilots at the pilot placement set $\bf{p}$, ${\bf{X}}_d ({\bf{p}})= diag\{[X(k_1),X(k_2),...,X(k_P)]\}$ denotes the transmitted signal matrix at $\bf{p}$, ${\bf{H}}_{vec}({\bf{p}})= [H(k_1,k_1),H(k_2,k_2),...,H(k_P,k_P)]^T$ denotes the channel responses at $\bf{p}$, ${\bf{d}} = {\bf{H}}_{ICI}({\bf p},:){\bf X}$ denotes the ICI at the pilot subcarriers, ${\bf{H}}_{ICI}({\bf p},:)$ denotes the rows of ${\bf H}_{ICI}$ at ${\bf p}$, and ${\bf{W}}({\bf{p}})$ denotes the AWGN at ${\bf p}$, respectively.

The theory of CS \cite{14} and \cite{15}  show that if ${\bf{b}}_x$ is $S$-sparse,  which is  satisfied in our system as aforementioned, then CS recover methods can  reconstruct ${\bf{b}}_x$ from ${\bf{Y}}({\bf{p}})$ successfully.
In this way, the task of estimating the high-mobility channel $\bf{H}$ in the frequency domain is converted to estimating the sparse dominant channel coefficients ${\bf{b}}_x$ in the delay-Doppler domain.

\subsection{ICI Mitigation}\label{ici}
In high-mobility environments, the transmitted pilots are distorted by the ICI coming from data and  AWGN as represented in Eq.~(\ref{eq8}), which highly affect the channel estimation performance.
In this paper, we adopt the ICI mitigation method proposed in \cite{93}.
Firstly, the high-mobility channels can be estimated by the proposed method.
As data is passed through the estimated channel, it provides an approximation of the data-induced ICI at the pilot subcarriers. Then, the estimated ICI can be subtracted at the pilot subcarriers. In this way, ICI caused by data is reduced and introduces better channel estimation performance. The process can be represented mathematically as:
\begin{equation}
{\bf{Y}}^{(q)} = {\bf{Y}}^{(q-1)} - \tilde{{\bf{H}}}^{(q-1)}{\bf{z}}^{(q-1)},
\end{equation}
where the superscript $q = 1,2,...$ denotes the iteration times. At each OFDM symbol, ${\bf{Y}}^{(0)}$ is the received  signal without ICI mitigation,
 $\tilde{{\bf{H}}}^{(q-1)}$ is the estimated channel in the  previous iteration, and ${\bf{z}}^{(q-1)}$ is the symbol constructed from the estimated data in the previous iteration with zeros at the pilot subcarriers.

In this manner,  the data-induced ICI at the pilot subcarriers can be reduced and get better system performance.
This process can be executed with more iteration times to further reduce the ICI but limited to a certain level due to the error propagation.
However, as the main topic of this work is the position-based channel estimator and pilot design, we do not consider the correct decision in this paper, and assume that the decision feedback equalizer is error-free to simplify the analysis.

\section{Coherence Optimized Pilot Design}

In this section, we first review some basis of CS and then formulate the pilot design problem. A joint pilot placement and pilot symbol design algorithm is proposed and discussed. Furthermore, the practical applicability of the proposed scheme is discussed.

\subsection{CS Fundamentals}
Compressed sensing is an innovative and revolutionary technique to reconstruct sparse signals accurately from a limited number of measurements.
Given an unknown signal ${\bf{x}}\in \mathbb{C}^m$, suppose ${\bf{x}}$ can be represented with a known dictionary matrix ${\bf{D}}\in {\mathbb{C}}^{m \times U}$ ($m<U$) and a vector ${\bf a} \in {\mathbb{C}}^{U}$, then we have $\bf{x}=\bf{D}\bf{a}$. Assume that $\bf{a}$ is $S$-sparse, i.e. $\|{\bf{a}}\|_{{{{\ell }_{0}}}}=S\ll U$.
CS considers the following problem
\begin{equation}
{\bf{y}}={\bf{P}}{\bf{x}} + {\bm{\eta}} =\bf{P}\bf{D}\bf{a} + {\bm{\eta}},
\end{equation}
in which ${\bf{P}}\in {\mathbb{C}}^{p \times m}$ denotes a known measurement matrix, ${\bf y} \in {\mathbb{C}}^{p}$ denotes the observed vector, and ${\bm{\eta}} \in  {\mathbb{C}}^{m}$ is the noise vector.
The objective of CS is to reconstruct $\bf{a}$ correctly based on the knowledge of $\bf y$, $\bf P$ and  $\bf{D}$.  Fundamental researches \cite{12} and \cite{14} indicate that if $\bf {PD}$ satisfies the restricted isometry property (RIP) \cite{15}, then $\bf a$ can be reconstructed correctly with CS reconstruction methods such as the basis pursuit (BP) \cite{17} and the orthogonal matching pursuit (OMP) \cite{18}.

\setcounter{TempEqCnt}{\value{equation}} 
\setcounter{equation}{23}                           
\begin{figure*}[!b]
\hrulefill
\begin{align}
{{\mu }_{\delta }}\left\{ {{\mathbf{X}}_{d}}(\mathbf{p})\mathbf{\Phi } \right\}=\frac{\sum\limits_{u\ne v}{\left( \left| \sum\limits_{{{k}_{i}}\in \mathbf{p}}{{{\left| X({{k}_{i}}) \right|}^{2}}\phi {{({{k}_{i}},u)}^{H}}\phi ({{k}_{i}},v)} \right|\ge \delta  \right)\cdot }\left| \sum\limits_{{{k}_{i}}\in \mathbf{p}}{{{\left| X({{k}_{i}}) \right|}^{2}}\phi {{({{k}_{i}},u)}^{H}}\phi ({{k}_{i}},v)} \right|}{\sum\limits_{u\ne v}{\left( \left| \sum\limits_{{{k}_{i}}\in \mathbf{p}}{{{\left| X({{k}_{i}}) \right|}^{2}}\phi {{({{k}_{i}},u)}^{H}}\phi ({{k}_{i}},v)} \right|\ge \delta  \right)}},
\label{ob1}
\end{align}
\end{figure*}
\setcounter{equation}{\value{TempEqCnt}}

\setcounter{TempEqCnt}{\value{equation}} 
\setcounter{equation}{25}                           
\begin{figure*}[!b]
\hrulefill
\begin{align}
{{\mu }_{\delta }}\left\{ {{\mathbf{X}}_{d}}(\mathbf{p})\mathbf{\Phi } \right\}=\frac{\sum\limits_{0\le u<v\le L(2M+1)-1}{\left( \left| \sum\limits_{t=1}^{T}{\sum\limits_{{{k}_{{{j}_{t}}}}\in {{\mathbf{s}}_{t}}}{{{E}_{t}}\cdot \phi {{({{k}_{{{j}_{t}}}},u)}^{H}}\phi ({{k}_{{{j}_{t}}}},v)}} \right|\ge \delta  \right)\cdot }\left| \sum\limits_{t=1}^{T}{\sum\limits_{{{k}_{{{j}_{t}}}}\in {{\mathbf{s}}_{t}}}{{{E}_{t}}\cdot \phi {{({{k}_{{{j}_{t}}}},u)}^{H}}\phi ({{k}_{{{j}_{t}}}},v)}} \right|}{\sum\limits_{0\le u<v\le L(2M+1)-1}{\left( \left| \sum\limits_{t=1}^{T}{\sum\limits_{{{k}_{{{j}_{t}}}}\in {{\mathbf{s}}_{t}}}{{{E}_{t}}\cdot \phi {{({{k}_{{{j}_{t}}}},u)}^{H}}\phi ({{k}_{{{j}_{t}}}},v)}} \right|\ge \delta  \right)}}.\label{eq6}
\end{align}
\end{figure*}
\setcounter{equation}{\value{TempEqCnt}}

\setcounter{TempEqCnt}{\value{equation}} 
\setcounter{equation}{26}                           
\begin{figure*}[!b]
\begin{align}
{{\mu }_{\delta }}\left\{ {{\mathbf{X}}_{d}}(\mathbf{p})\mathbf{\Phi }_x \right\}=\frac{\sum\limits_{u<v~\text{and}~u,v\in {{\mathbf{\Phi }}_{x}}}{\left( \left| \sum\limits_{t=1}^{T}{\sum\limits_{{{k}_{{{j}_{t}}}}\in {{\mathbf{s}}_{t}}}{{{E}_{t}}\cdot \phi {{({{k}_{{{j}_{t}}}},u)}^{H}}\phi ({{k}_{{{j}_{t}}}},v)}} \right|\ge \delta  \right)\cdot }\left| \sum\limits_{t=1}^{T}{\sum\limits_{{{k}_{{{j}_{t}}}}\in {{\mathbf{s}}_{t}}}{{{E}_{t}}\cdot \phi {{({{k}_{{{j}_{t}}}},u)}^{H}}\phi ({{k}_{{{j}_{t}}}},v)}} \right|}{\sum\limits_{u<v~\text{and}~u,v\in {{\mathbf{\Phi }}_{x}}}{\left( \left| \sum\limits_{t=1}^{T}{\sum\limits_{{{k}_{{{j}_{t}}}}\in {{\mathbf{s}}_{t}}}{{{E}_{t}}\cdot \phi {{({{k}_{{{j}_{t}}}},u)}^{H}}\phi ({{k}_{{{j}_{t}}}},v)}} \right|\ge \delta  \right)}}.\label{eq25}
\end{align}
\end{figure*}
\setcounter{equation}{\value{TempEqCnt}}

\setcounter{TempEqCnt}{\value{equation}} 
\setcounter{equation}{27}                           
\begin{figure*}[!b]
\begin{align}
{{\mu }_{\delta }}\left\{ {{\mathbf{X}}_{d}}(\mathbf{p})\mathbf{\Phi }_x \right\}=\frac{\sum\limits_{u<v~\text{and}~u,v\in {{\mathbf{\Phi }}_{x}}}{\left( \left| \sum\limits_{t=1}^{T}{\sum\limits_{{{k}_{{{j}_{t}}}}\in {{\mathbf{s}}_{t}}}{{{E}_{t}}\cdot {{e}^{-j\frac{2\pi }{W}(v-u){{k}_{{{j}_{t}}}}}}}} \right|\ge \delta  \right)\cdot }\left| \sum\limits_{t=1}^{T}{\sum\limits_{{{k}_{{{j}_{t}}}}\in {{\mathbf{s}}_{t}}}{{{E}_{t}}\cdot {{e}^{-j\frac{2\pi }{W}(v-u){{k}_{{{j}_{t}}}}}}}} \right|}{\sum\limits_{u<v~\text{and}~u,v\in {{\mathbf{\Phi }}_{x}}}{\left( \left| \sum\limits_{t=1}^{T}{\sum\limits_{{{k}_{{{j}_{t}}}}\in {{\mathbf{s}}_{t}}}{{{E}_{t}}\cdot {{e}^{-j\frac{2\pi }{W}(v-u){{k}_{{{j}_{t}}}}}}}} \right|\ge \delta  \right)}}.\label{eq26}
\end{align}
\end{figure*}
\setcounter{equation}{\value{TempEqCnt}}

\setcounter{TempEqCnt}{\value{equation}} 
\setcounter{equation}{28}                           
\begin{figure*}[!b]
\begin{align}
{{\mu }_{\delta }}\left\{ {{\mathbf{X}}_{d}}(\mathbf{p})\mathbf{\Phi }_x \right\}=\frac{\sum\limits_{1\le z\le L-1}{\left( \left| \sum\limits_{t=1}^{T}{\sum\limits_{{{k}_{{{j}_{t}}}}\in {{\mathbf{s}}_{t}}}{{{E}_{t}}\cdot {{e}^{-j\frac{2\pi }{W}z{{k}_{{{j}_{t}}}}}}}} \right|\ge \delta  \right)\cdot }\left| \sum\limits_{t=1}^{T}{\sum\limits_{{{k}_{{{j}_{t}}}}\in {{\mathbf{s}}_{t}}}{{{E}_{t}}\cdot {{e}^{-j\frac{2\pi }{W}z{{k}_{{{j}_{t}}}}}}}} \right|}{\sum\limits_{1\le z\le L-1}{\left( \left| \sum\limits_{t=1}^{T}{\sum\limits_{{{k}_{{{j}_{t}}}}\in {{\mathbf{s}}_{t}}}{{{E}_{t}}\cdot {{e}^{-j\frac{2\pi }{W}z{{k}_{{{j}_{t}}}}}}}} \right|\ge \delta  \right)}}.\label{27}
\end{align}
\end{figure*}
\setcounter{equation}{\value{TempEqCnt}}



To improve the CS performance, in this paper, we consider the average coherence proposed in \cite{13}. It has been proved in \cite{13} that the average coherence reflects the actual CS behavior rather than the mutual coherence \cite{12} for considering the average performance. The definition is given as follows.

\begin{definition}[Average Coherence \cite{13}]
For a matrix $\bf{M}$ with the $i$-th column of ${\bf{d}}_i$, its average coherence is defined as the average of all absolute inner products between the different normalized columns in $\bf{M}$ that are beyond a threshold $\delta$, where $0<\delta<1$. Put formally
\begin{equation}
\mu _\delta \{ {\bf{M}}\}  = \frac{{\sum\limits_{i \ne j} {\left( {\left| {g_{ij} } \right| \ge \delta} \right) \cdot \left| {g_{ij} } \right|} }}{{\sum\limits_{i \ne j} {\left( {\left| {g_{ij} } \right| \ge \delta} \right)} }},\label{eq3}
\end{equation}
where $g_{ij} = \tilde{\bf{ d}}^H_i \tilde{\bf{ d}}_j$, $\tilde {\bf{ d}}_i ={\bf{ d}}_i/\|{\bf{ d}}_i\|_{\ell_2} $, and
the operator is defined as
\begin{equation}
(x\geq y)=\left\{\begin{matrix}
  &1,~&x\geq y ,\\
  &0,~&x<y.
\end{matrix}\right.
\end{equation}
\end{definition}

Previous researches \cite{12} and \cite{30}  established that BP and orthogonal greedy algorithms (OGA) (including OMP) can recover $\bf a$ correctly provided that the following theorem is satisfied.

\begin{theorem}[\cite{12}]
For a dictionary matrix $\bf{D}$ and a measurement matrix $\bf{P}$, assume that $\bf{PD}$ satisfies  RIP. If ${\bf {y}   =  { \bf{ P} } \bf{x}   =  \bf{P }\bf{D} \bf{a} }$ satisfies
\begin{equation}
S={\left\| {{{\bf{a}}}} \right\|_{\ell_0}} < \frac{1}{2}\left( {1 + \frac{1}{{{\mu_\delta}\left\{ {\bf{PD}} \right\}}}} \right),\label{eq7}
\end{equation}
then a)  $\bf{a}$ is the unique sparsest  representation of $\bf{x}$; b) the deviation of the reconstructed ${\hat{\bf{a}}}$ from $\bf{a}$ by BP or OGA can be bounded by
\begin{equation}
\|{\hat{\bf{a}}} - {\bf{a}}\|^2_{{\ell }_2}\leq \frac{\epsilon^2}{1-\mu_\delta\{{\bf{PD}}\}(2S-1)},
\end{equation}
for some constant $\epsilon>0$.
\end{theorem}

It is easy to find that a smaller $\mu_\delta \{\bf{PD}\}$ will lead to a lower reconstruction error bound which results in a more accurate recovery of $\bf a$. Furthermore, Theorem $1$ implies that if $\bf{P}$ is designed with a fixed $\bf{D}$ such that ${{\mu_\delta}\left\{ {\bf{PD}} \right\}}$ is as small as possible, a large number of candidate signals are able to reside under the umbrella of successful CS behavior and lead to a better performance.

\setcounter{TempEqCnt}{\value{equation}} 
\setcounter{equation}{29}                           
\begin{figure*}[!b]
\hrulefill
\begin{align}
{\bf{ X}}^*_d&=\underset{\left|{\bf{{X}}}_d\right|,{\bf{p}}}{{\mathop{\arg\min} }}\,{{\mu_\delta }}\left\{ {{\mathbf{X}}}_d({\bf{p}}){{\mathbf{\Phi }}}_x \right\}, \label{eq28}\\
&=\underset{{\bf{s}}_t,E_t}{{\mathop{\arg \min} }}\,
\frac{\sum\limits_{1\le z\le L-1}{\left( \left| \sum\limits_{t=1}^{T}{\sum\limits_{{{k}_{{{j}_{t}}}}\in {{\mathbf{s}}_{t}}}{{{E}_{t}}\cdot {{e}^{-j\frac{2\pi }{W}z{{k}_{{{j}_{t}}}}}}}} \right|\ge \delta  \right)\cdot }\left| \sum\limits_{t=1}^{T}{\sum\limits_{{{k}_{{{j}_{t}}}}\in {{\mathbf{s}}_{t}}}{{{E}_{t}}\cdot {{e}^{-j\frac{2\pi }{W}z{{k}_{{{j}_{t}}}}}}}} \right|}{\sum\limits_{1\le z\le L-1}{\left( \left| \sum\limits_{t=1}^{T}{\sum\limits_{{{k}_{{{j}_{t}}}}\in {{\mathbf{s}}_{t}}}{{{E}_{t}}\cdot {{e}^{-j\frac{2\pi }{W}z{{k}_{{{j}_{t}}}}}}}} \right|\ge \delta  \right)}}.\label{eq29}
\end{align}
\end{figure*}
\setcounter{equation}{\value{TempEqCnt}}

\setcounter{TempEqCnt}{\value{equation}} 
\setcounter{equation}{31}                           
\begin{figure*}[!b]
\hrulefill
\begin{align}
   Pr \{\mu_\delta \{\mathbf{X}_{m}^{*}{{\mathbf{\Phi }}_{x}}\}<\mu_\delta  \{{{\mathbf{X}}_{m}}{{\mathbf{\Phi }}_{x}}\}\}>Pr \{\mu_\delta  \{{{\mathbf{X}}_{m}}{{\mathbf{\Phi }}_{x}}\}<\mu_\delta  \{\mathbf{X}_{m}^{*}{{\mathbf{\Phi }}_{x}}\}\}, \label{eq31}\\
 Pr \{\mu_\delta \{\mathbf{X}_{m}^{*}{{\mathbf{\Phi }}_{x}}\}<\mu_\delta  \{{{\hat{\mathbf{{X}}}}_{m}}{{\mathbf{\Phi }}_{x}}\}\}>Pr \{\mu_\delta  \{{{\hat{\mathbf{{X}}}}_{m}}{{\mathbf{\Phi }}_{x}}\}<\mu_\delta  \{\mathbf{X}_{m}^{*}{{\mathbf{\Phi }}_{x}}\}\}\label{eq32}.
\end{align}
\end{figure*}
\setcounter{equation}{\value{TempEqCnt}}

\subsection{Problem Formulation}
As we have already known that a lower $\mu_\delta$ leads to a better CS performance,  we are going to reduce  ${{\mu_\delta }\left\{ {\bf{X}}_{d} ({\bf{p}}){\bf{\Phi}}_x \right\}}$ in our system to get better  performance. 
In this paper, both the pilot placement and pilot symbol of the transmitted pilot matrix ${\bf{X}}_{d}(\bf{p})$ are considered with the  dominant channel model dictionary ${\bf{\Phi}}_x$.

Let us start from the objective of minimizing  ${{\mu_\delta }\left\{ {\bf{X}}_{d} (\bf{p}){\bf{\Phi}} \right\}}$. This pilot design problem can be formulated as follows
\begin{equation}
\underset{\left|{\bf{{X}}}_d\right|,{\bf{p}}}{\mathop{\min }}\,{{\mu_\delta }}\left\{ {{\mathbf{X}}}_{d}(\bf{p}){{\mathbf{\Phi }}} \right\},
\end{equation}
where $| {\bf{X}}_d|$ denotes the pilot symbols in ${\bf{X}}_{d}(\bf{p})$ and ${\bf{p}}$ denotes the pilot placement set. According to Definition 2, the objective function can be represented as Eq. (\ref{ob1}),
 where ${\phi }({{k}_{i}},u)$ is the entry of  ${\bf{\Phi}}$ and $0\le u < v\le L(2M+1)-1$.

Suppose that all pilots and data are modulated symbols, and there are $T$ pilot powers levels corresponding to $T$ pilot  placement subsets $\{{\mathbf{s}}_{t}\} _{t \in T}$. Then we have $\bigcup\limits_{t=1}^{T}{{{\mathbf{s}}_{t}}=\mathbf{p}}$ and define the pilot power as
\addtocounter{equation}{1}
\begin{equation}
{{E}_{t}} \triangleq {{\left| X({{k}_{j_t}}) \right|}^{2}},~k_{j_t}\in {{\bf{s}}_{t}}, \label{eq21}
\end{equation}
for ${j_t}\in\{1,2,...,P\}$ and $t=1,2,...,T$.
By taking pilot powers into consideration, (\ref{ob1}) can be represented as Eq. (\ref{eq6}).

Furthermore, we consider the proposed position-based channel model. When the HST moves to a certain position $\alpha$, the high-mobility channels can be modeled by ${\bf{b}}_x$ corresponding to $f_d$. Therefore, we only need to consider the property of the columns in ${\bf{\Phi}}_x$. The columns in ${\bf{\Phi}}_x$ can be represented as  ${{\phi}}(:,u)$ and $u\in [L(M+x),L(M+x)+L-1]$.
Then (\ref{eq6}) with $\mathbf{\Phi }_x $ can be further represented as Eq. (\ref{eq25}).
By taking the expression of ${\bf{\Phi}}_x$ into consideration, we have Eq. (\ref{eq26}).
Denote $z = v-u$. Then the objective function is simplified as Eq. (\ref{27}).
So far, the optimization variable has be simplified from a $P\times L(2M+1)$ matrix ${{\mathbf{X}}_{d}}(\mathbf{p})\mathbf{\Phi }$ to a $P\times L$ matrix ${{\mathbf{X}}_{d}}(\mathbf{p})\mathbf{\Phi }_x$, and the number of calculations has been reduced from $\binom{L(2M+1)}{2}$ to $(2M+1)\binom{L}{2}$.

Finally, we formulate the pilot design problem as a joint optimization problem as Eq. (\ref{eq29}).

\begin{algorithm}[!t]\small
 \caption{{\bf{:}} Joint Pilot Placement and Pilot Symbol Design Algorithm}
  \begin{algorithmic}[1]
\raggedright
\Require Initial pilot  ${\bf{X}}_0$ with the pilot placement ${\bf{p}}_0$ and the pilot symbol ${\bf{x}}_0$.
\Ensure Optimal pilot ${\bf{X}}^*_d$.

\State{\bf{Initialization}}: Set $\hat{\bf{X}}_0 = {\bf{X}}_0$; set  $M$, set  $Iter = M\times P$; set pilot powers $E_1,E_2,...,E_T$; set ${\bf{I}}[0] = {\bf{0}}$, ${I}[0,0] = 1$; set $\kappa = 0, \iota = 0$.

\For{$ n = 0, 1,..., M - 1$}
    \For{$ k = 0, 1,..., P - 1$}
        \State $m \Leftarrow n\times P + k$;\\
~~~~~$\diamond$ \underline{{Generate and update}}:
        \State generate $\tilde{\bf{p}}_m$ with operator $\tilde{\bf{p}}_m ({\bf{p}}_m)$;
        \If {$\mu_\delta\{{\bf{X}}_m({{\tilde{\bf{p}}}_m}){\bf{\Phi}}_x\} < \mu_\delta\{{\bf{X}}_m({{{\bf{p}}}_m}){\bf{\Phi}}_x\}$}
         \State select  pilot symbol power $E_t$ to $\min\mu_\delta$;
         \State update ${\bf{x}}_m$ and ${\bf{x}}_{m+1} \Leftarrow {\bf{x}}_m$; ${\bf{p}}_{m+1} \Leftarrow {{\tilde{\bf{p}}}_m}$;
        \Else
         \State select  pilot symbol power $E_t$ to $\min\mu_\delta$;
         \State update ${\bf{x}}_m$ and ${\bf{x}}_{m+1} \Leftarrow {\bf{x}}_m$; ${{{\bf{p}}}_{m+1}} \Leftarrow {{{\bf{p}}}_m}$;
        \EndIf
        \State update ${\bf{X}}_{m+1}$ with ${\bf{p}}_{m+1}$ and ${\bf{x}}_{m+1}$; $\kappa \Leftarrow m+1$;\\
~~~~~$\diamond$ \underline{{Update state occupation probabilities}}:
        \State ${\bf{I}}[m+1] \Leftarrow {\bf{I}}[m] + \eta[m+1]({\bf{D}}[m+1] - {\bf{I}}[m])$, with $\eta[m]= 1/m$;
        \If {${{I}}[m+1,\kappa] > {{I}}[m+1,\iota]$}
            \State $\hat{\bf{X}}_{m+1} \Leftarrow {\bf{X}}_{m+1}$; $\iota \Leftarrow \kappa$;
        \Else
            \State $\hat{\bf{X}}_{m+1} \Leftarrow \hat{\bf{X}}_{m}$;
        \EndIf
    \EndFor ~(k)
\EndFor ~(n)
\end{algorithmic}
\end{algorithm}

\subsection{Joint Pilot Placement and Pilot Symbol Design Algorithm}

An intuitive method to find the global optimal solution is to perform exhaustive search over ${{T}^{P}}\binom{K}{P}$ combinations. This method however is impractical for an OFDM system with large $K$ due to huge computational complexity. In this subsection, we treat this optimization problem as a discrete stochastic optimization problem \cite{40} and propose an iterative algorithm to solve it.
%

The key idea of our algorithm is to generate a sequence of candidate pilot matrices, where each new candidate is obtained from the previous one by taking a step in a  direction towards the global optimum.
The details are given in Algorithm 1.
Define ${\bf{p}}_m$, $\tilde{\bf{p}}_m$ and $\hat{\bf{p}}_m$ as different pilot placement sets at the $m$-th iteration. $M$ is the number of  pilot placement sets, and $Iter = M \times P$ denotes the total iteration times.
At each iteration, the algorithm updates the probability vector ${\bf{I}}[m] = (I[m,1],I[m,2],...,I[m,MP])^T$, which represents the state occupation probabilities of the generated pilot matrices with entries ${{I}}[m,\kappa] \in [0,1]$ , and $\sum_{\kappa}{{I}}[m,\kappa] =1$.
${\bf{D}}[m] \in \mathbb{R}^{MP\times 1}$ is defined as a zero vector except for its $m$-th entry to be 1.

Algorithm 1 starts with an initial pilot ${\bf{X}}_0$ with a random pilot placement set ${\bf{p}}_0$ and a random pilot symbol vector ${\bf{x}}_0$.
In the \emph{Generate and update} step, $\tilde{\bf{p}}_m$ is obtained uniformly with the operator $\tilde{\bf{p}}_m ({\bf{p}}_m)$. At the $m$-th iteration, the $k$-th pilot subcarrier of ${\bf{p}}_m$ is replaced with a random subcarrier which is not included in  ${\bf{p}}_m$ and then gets $\tilde{\bf{p}}_m$.
 Then we compare ${{\tilde{\bf{p}}}_m}$ with ${{{\bf{p}}}_m}$ and select the better one to move a step.
Furthermore, we minimize the $\mu_\delta$ and select the best symbol power to update.
In the \emph{Update state occupation probabilities} step, ${\bf{I}}[m+1]$ is updated based on the previous ${\bf{I}}[m]$ with the decreasing step size $\eta[m] = 1/m$.  $\eta[m]$ avoids the proposed algorithm moving away from a promising point unless there was a strong evidence that this move will result in an improvement, which makes Algorithm 1 more progressive and conservative with increasing iterations.

\subsection{Global Convergence Property}
The sequence $\{{\bf{X}}_m\}$ generated by the proposed algorithm  is a Markov chain which in general cannot converge to a fixed point and may visit each entry infinitely often. In this subsection, we show that the sequence $\{\hat{\bf{X}}_m\}$ surely converges to the global optimal solution ${\bf{X}}^*$ under certain conditions.
The sufficient conditions for Algorithm 1 to converge to ${\bf{X}}^*$ are given as ({\ref{eq31}}) and ({\ref{eq32}}) \cite{41}.
For generated solutions ${\bf{X}}_m \neq {\bf{X}}^*$ and $\hat{\bf{X}}_m \neq {\bf{X}}^*$, if (\ref{eq31}) and (\ref{eq32}) are satisfied, then \cite{41} proves that $\{{\bf{X}}_m\}$ is a homogeneous aperiodic and irreducible Markov chain in its state space. Moreover, as  $\{{\bf{X}}_m\}$ spends much more efforts in ${\bf{X}}^*$ than others, $\{\hat{\bf{X}}_m\}$ surely converges to ${\bf{X}}^*$.

Condition (\ref{eq31}) ensures that it is more probable for $\{{\bf{X}}_m\}$ to move into a state corresponding to ${\bf{X}}^*$ than others. Condition (\ref{eq32}) ensures that once $\{{\bf{X}}_m\}$ is in a state that not corresponding to ${\bf{X}}^*$, it is more probable for $\{{\bf{X}}_m\}$ to move into a state that corresponding to ${\bf{X}}^*$ than others.
Therefore, Algorithm 1 is a globally convergent algorithm which spends most of time at the global optimum.
In addition, the property of spending more time at the global optimum than any other solution is called the attraction property of algorithms \cite{43}. The  attraction property shows that Algorithm 1 is efficient.

\subsection{Complexity Analysis}
The computational complexity is determined in terms of the number of the complex multiplications needed.
The proposed algorithm consists of three steps: \emph{Initialization}, \emph{Generate and update} and \emph{Update state occupation probabilities}.
In the \emph{Initialization} step, matrices and parameters are pre-computed and stored in the memory, so its computational complexity can be omitted.
The \emph{Generate and update} step computes the objective function with different pilots, which requires $M(T+2){P^2L(L-1)}/{2}$ complex multiplications.
Note that the proposed position-based channel estimator highly reduces the multiplications of this step from $M(T+2)P^2L(2M+1)[L(2M+1)-1]/2$, especially for a large $M$ corresponding to a high Doppler shift.
Regarding to the \emph{Update state occupation probabilities} step, this update requires $M^2P^2$ multiplications. Therefore, Algorithm 1 requires $M(T+2){P^2L(L-1)}/{2} + M^2P^2$ complex multiplications in total.
In a practical system, $M$, $L$ and $T$ are constant parameters and much smaller than $N$. Since the pilot number $P$ is much smaller than $N$ in practice, the computational complexity of Algorithm 1 is much lower than  ${\mathcal{O}}(N^2)$.
In contrast, the complexity of \cite{20} is  ${\mathcal{O}}(N^3)$.
Furthermore, as the needed system parameters can be estimated in advance, Algorithm 1 is an off-line process and thus its complexity can be ignored in the practical system.

\subsection{Practical Applicability}
In this subsection, we briefly discuss the applicability of the proposed scheme in a practical HST system. As aforementioned in Section II-A, the BSs are connected with optical fibers and  share the instant speed and position information of the HST, which are  supported by the GPS. In a practical system, system parameters (such as  $\tau _{\max }$, $f_{d_{\max }}$ and etc.) can be collected in advance. Thus the optimal pilots  (including the pilot placement and pilot symbol) for different Doppler shifts $f_d$ (or HST positions $\alpha$) can be pre-designed with Algorithm 1 by selecting corresponding ${\bf \Phi}_x$, and then store into a pilot codebook, which is an off-line process. The relationships between $x$, $f_d$ and $\alpha$ are given as Eq. (\ref{eq12}) and Eq. (\ref{eq13}), respectively. This pilot codebook is equipped at both the BS and the HST.

When the HST  runs, the BS gets the instant speed and position information of the HST from the GPS and then calculates the instant $f_d$. At the beginning of each OFDM symbol, the BS selects the optimal pilot from the codebook according to $f_d$ and transmits it to estimate the channels. This transmitted pilot is also known at the HST for checking the same codebook.
Note that we assume that $f_d$ is constant during one OFDM symbol. Thus, the selected pilot is optimal during each OFDM symbol.
In this way, the proposed scheme can be well used in current HST systems without adding too much complexity.
An example of the designed pilot codebook is given in the simulation results.

\begin{table}[!t]
\centering
\caption{\label{table1} HST COMMUNICATION SYSTEM PARAMETERS} \label{tb1}
\renewcommand\arraystretch{1.1}
\begin{tabular}{c c c} \toprule
{Parameters} & {Variables}&{Values} \\  \toprule
BS power range  & $R$     &    $ 1200$~m      \\
Distance between BSs  & $Ds$     &    $1000$~m      \\
Max distance of BS to railway  & $D_{max}$    &     $1200$~m \\
Direct distance of BS to railway  & $D_0$    &     $50$~m     \\
Carrier frequency & $f_c$ & $2.35$~GHz \\
Train speed & $v$    &    $300$-$500$~km/h       \\
Light speed & $c$    &    $3\times 10^8$~m/s     \\ \bottomrule
\end{tabular}
\end{table}

\section{Simulation Results}

In this section, under the high-mobility environment, we illustrate the performances of
 the proposed pilot design method using two typical compressed channel estimators,  BP \cite{17} and OMP \cite{18}.
The mean square error (MSE) and the bit error rate (BER) performances are considered versus the the signal to noise ratio (SNR) and the HST position.
In addition, the performances of the conventional LS and the linear minimum mean square error (LMMSE) \cite{6} estimators are also considered. Furthermore, we give an example of the designed pilot codebook.



Here we consider an OFDM system in the HST communication system shown as Fig.~1. The parameters of the HST communication system are given in Table \ref{tb1}, according to the D2a scenario of WINNER II channel model \cite{44}.
Assume that there are $512$ subcarriers and $12.5\%$ are pilot subcarriers.
The bandwidth is $5$MHz, the packet duration is $T_d = 0.675$ms and the carrier frequency $f_c=2.35$GHz, according to \cite{9}. Data and pilots are modulated with $16$-QAM.
The additive noise is a Gaussian and white random process. The maximum delay spread is $\tau_{max} = 5\mu s$ and the maximum Doppler shift is $f_{d_{max}} = 1.088$KHz, which means that the maximum speed of the HST is 500km/h. The channel has $L = 26$ taps, however, only $6$ taps are nonzero and their positions are randomly generated.
The ICI mitigation is operated as mentioned in Section \ref{ici}.

%

\subsection{Doppler Shift versus HST Position}
\begin{figure}[!t]
\includegraphics[width=3.6in]{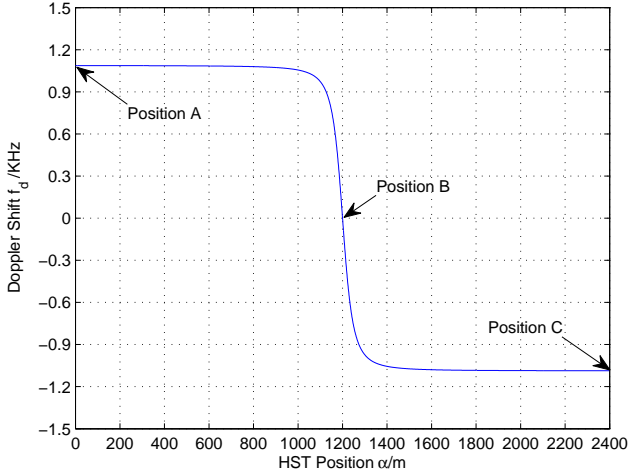}
\caption{Doppler shift versus HST position.}
\label{fig4}
\end{figure}

\reffig{fig4} shows the Doppler shift of the proposed position-based high-mobility OFDM system versus the HST position. The speed of HST is 500km/h. In \reffig{fig4}, the Doppler shift changes from $f_{d_{max}}$ to $-f_{d_{max}}$ corresponding to the HST position $\alpha$. $\alpha$ is defined as the distance between the HST and $A$, and $\alpha = 0$ at $A$.
We can find that HST suffers from large Doppler shift at most of the time, and $f_d$  changes sharply near $B$.  



\begin{figure}[!t]
\centering
\includegraphics[width=3.6in]{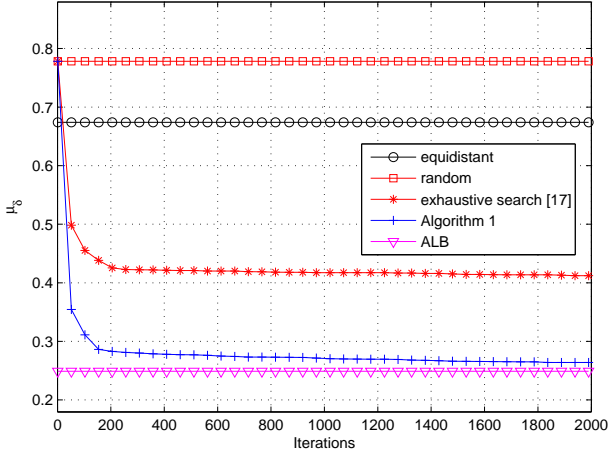}
\caption{Average (500 runs) of average coherence $\mu_\delta\{{\bf{X}}_d({{{\bf{p}}}}){\bf{\Phi}}_x\}$ with different pilot design methods in an OFDM system with 500km/h. }\label{fig6}
\end{figure}

\begin{figure}[!t]
\centering
\includegraphics[width=3.6in]{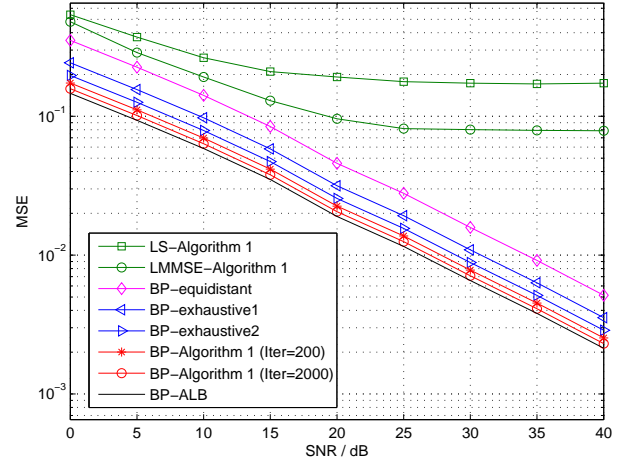}
\caption{MSE performances of liner estimators and BP estimators  with different pilots in an OFDM system at position $A$ with 500km/h. }\label{fig7}
\end{figure}

\begin{figure}[!t]
\centering
\includegraphics[width=3.6in]{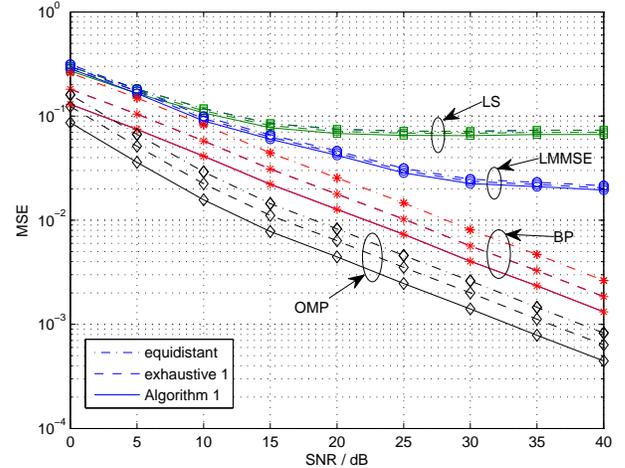}
\caption{MSE performances of different estimators with different pilots in an OFDM system at position $A$ with 325 km/h. }\label{fig8}
\end{figure}


\subsection{Average Coherence versus Iterations}
In \reffig{fig6}, we consider 500 channel realizations and gives the average coherence $\mu_\delta$ performance.
The equidistant method is the pilot  with the equidistant pilot placement and random pilot symbols,  which is claimed in \cite{42} as the optimal pilot placement to the doubly selective channels.
We also plot the the exhaustive search method in \cite{20} whose main idea is to do an exhaustive search from a designed optimal pilot subset.  The Algorithm 1 with $Iter =  1\times 10^4$ is given to show the approximate lower bound (ALB) of Algorithm 1, which means that the performance improves extremely little by increasing iterations.
As can be seen, Algorithm 1 converges fast before $Iter = 200$ and then converges to  its ALB slowly.
Note that the pilot placement iteration times is $M = \frac{Iter}{P} \approx 3$ at $Iter = 200$, which means that Algorithm 1 calculates 3 set of pilots and then gets the optimal pilot.
Considering the tradeoff of the computational complexity and the estimation performance, we select $Iter=200$ in the following simulations.


\subsection{MSE versus SNR}

\reffig{fig7} illustrates the comparison of the MSE performances of different estimators with different pilots with 500km/h at position $A$. The ICI mitigation is operated with $q=2$. The exhaustive 1 method is the method in \cite{20} with $200$ iterations, and the exhaustive 2 method denotes the same method with $2\times 10^4$ iterations. The number of iterations of Algorithm 1 is set to be $200$ and $2000$.
It can be observed that the BP channel estimators significantly improve the performance by utilizing the inherent sparsity of the high-mobility channels.
As expected, Algorithm 1 gets better performance than others.
Furthermore, we notice that LS and LMMSE need more pilots to obtain better CSI, while the proposed scheme  performs well and saves spectrum resources.
In addition, BP with the ALB pilot  is given to show the convergence tendency of Algorithm 1. It can be seen that BP-Algorithm 1 converges to BP-ALB with increasing iterations. 
From the curves, it is possible to observe that Algorithm 1 with $Iter = 200$ is enough for the practical system.

\begin{figure}[!t]
\centering
\includegraphics[width=3.6in]{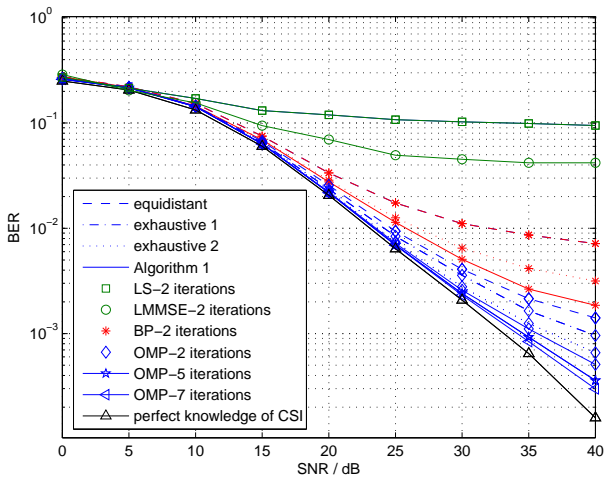}
\caption{BER performances of different estimators with different pilots and ICI mitigation iterations in an OFDM system at position $A$ with 500km/h.}\label{fig9}
\end{figure}

\begin{figure}[!t]
\centering
\includegraphics[width=3.6in]{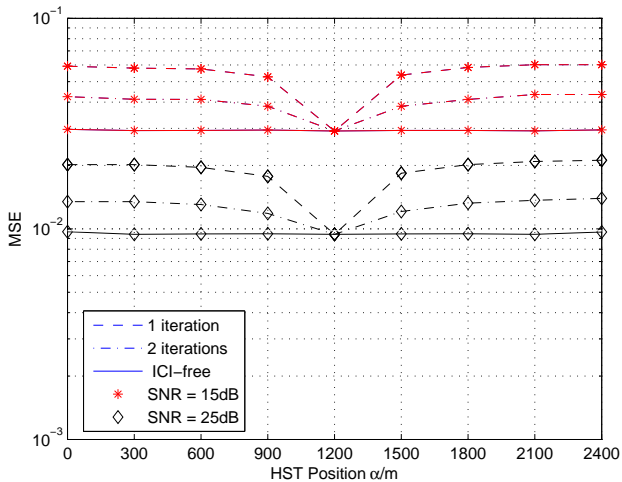}
\caption{MSE performances of BP estimators with Algorithm 1 versus  HST positions in an OFDM system at 500km/h.}\label{fig10}
\end{figure}

\begin{figure}[!t]
\centering
\includegraphics[width=3.7in]{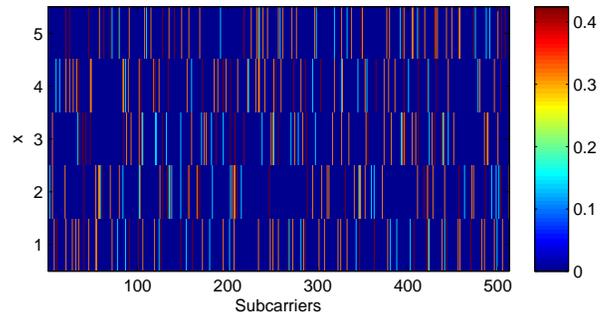}
\caption{A pilot codebook designed by Algorithm 1. }\label{fig11}
\end{figure}

\reffig{fig8} depicts the comparison of the MSE performances of different estimators versus  SNR at position $A$ with $325$km/h, which means $f_{d} = 0.707$KHz. The ICI mitigation is operated with $q = 2$ and Algorithm 1 is operated with $Iter = 200$. As can be seen, with Algorithm 1, BP and OMP get better performances comparing with other pilots.
On the other hand, LS and LMMSE with different pilots are also given in this figure. We notice that Algorithm 1 has little impact on linear estimators for their not utilizing the coherence of CS.

\subsection{BER versus SNR}

\reffig{fig9} shows the BER performances versus SNR in the given high-mobility environment at position $A$ with $500$km/h. As a reference, we plot the BER performance under the perfect knowledge of CSI with Algorithm 1, which means that $\bf{H}$ in Eq. (\ref{eq0}) is available at the receiver and employed with the zero-forcing (ZF) equalizer.
In this figure, we compare different estimators with the pilots designed by the equidistant method, the exhaustive 1 method, the exhaustive 2 method and Algorithm 1 ($Iter = 200$). ICI mitigation with 2, 5 and 7 iterations are considered to show its performance tendency. OMP-5 and OMP-7 are both operated with Algorithm 1.
As can be seen, BP and OMP with Algorithm 1 get significant improvements and are close to the  perfect knowledge of CSI, particularly in regions of low and moderate SNR. This is mainly because that, at low SNR, the noise is dominant with respect to the ICI. We also notice that the ICI mitigation gain is limited with increasing iterations due to  error propagation.


\subsection{MSE versus HST Position}

\reffig{fig10} presents the MSE performances of BP estimators versus the HST position at  SNR = $15$dB and SNR = $25$dB, in which the Doppler  shift $f_d$ is corresponding to the HST position $\alpha$ shown as \reffig{fig4}. All estimators are considered with the pilot designed  by Algorithm 1 ($Iter = 200$).
 The train speed is fixed as 500km/h, and $f_d$ changes from $f_{d_{max}}$  to $-f_{d_{max}}$ with the  HST moves from $A$ to $C$ shown as \reffig{fig4}.  As a reference, the performance with ICI-free are also included, in which ``ICI-free" means the data are set as zero and pilots are free of data-ICI.
 We notice that the MSE performances change with the HST position and get the best at $B$.
This is reasonable because $f_d$ is largest at $A$ and $C$ but zero at $B$.
This figure shows that the HST suffers from large Doppler shift at most of the time. However, the MSE performances improve rapidly near B with decreasing Doppler shift.
It can be seen that the proposed channel model well reflects the relationship between the Doppler shift and the HST position.

\subsection{A Pilot Codebook for Practical Use}

\reffig{fig11}  gives a pilot codebook designed by Algorithm 1 ($Iter = 200$) of the given system, which shows the optimal pilot sets (including the pilot placement and pilot symbol) according to the Doppler shift $f_d$ (or HST position $\alpha$). $x$ denotes the index of the pilot set, its relationship with $f_d$ and $\alpha$ are given as Eq. (\ref{eq12}) and Eq. (\ref{eq13}), respectively.
The pilot symbol powers are presented as different colors.
Furthermore, data are set to be zero for stressing the pilot placement.
When the HST position changes, the optimal pilot changes according to the instant $f_d$.
According to different $f_d$, the BS and RS  calculate $x$ with Eq. (\ref{eq12}) and select the optimal pilot set from the codebook.
For example, $x = 5$ is selected for $f_d \in[0.7407, 1.088]$KHz (near  $A$), and $x = 1$ is selected for  $f_d \in[-1.088, -0.7407]$KHz (near  $C$), respectively. Meanwhile, as $f_d$ changes rapidly when the HST passing $B$, pilot sets $x = 4,3,2$ are selected in sequence.
From \reffig{fig4}, we notice that the HST suffers from high Doppler shift at most of the time while passing through a cell. Thus, we do not need to change the pilot set frequently while the HST runs (except for the positions near $B$), which highly reduce the system complexity.

\section{Conclusion}
%

In this paper, we  presented a new position-based compressed channel estimation method for high-mobility OFDM systems. 
The estimation complexity is reduced by the proposed channel model by utilizing the  position information. The pilot symbol and the placement are jointly designed by the proposed algorithm to minimize the system average coherence.
Simulation results demonstrate that the proposed method achieves better performances than existing channel estimation methods over high-mobility channels.
Furthermore, with a pre-designed pilot codebook, the proposed scheme is feasible for  many current wireless OFDM communication systems.


\begin{thebibliography}{40}
\bibitem{1}
P. Schniter, ``Low-complexity equalization of OFDM in doubly selective channels," \emph{IEEE Transactions on Signal Processing,} vol. 52, no. 4, pp. 1002-1011, April 2004.

\bibitem{5}
O. B. Karimi, J. Liu, and C. Wang, ``Seamless wireless connectivity for multimedia services in high speed trains," \emph{IEEE Journal on Selected Areas in Communications,} vol. 30, no. 4, pp. 729-739, May 2012.

\bibitem{2}
W. U. Bajwa, A. M. Sayeed, and R. Nowak. ``Sparse multipath channels: modeling and estimation," \emph{IEEE Digital Signal Processing Education Workshop,} pp. 320-325, Jan. 2009.

\bibitem{3}
W. U. Bajwa, J. Haupt, A. M. Sayeed, and R. Nowak, ``Compressed channel sensing: a new approach to estimating sparse multipath channels," \emph{Proceedings of the IEEE}, vol. 98, no. 6, pp. 1058-1076, June 2010.

\bibitem{4}
W. U. Bajwa, A. M. Sayeed, and R. Nowak, ``Learning sparse doubly-selective channels," \emph{46th Annual Allerton Conference on Communication, Control and Computing,} pp. 575-582, Sept. 2008.

\bibitem{90}
S. Sung and D. Brady, ``Spectral spatial equalization for OFDM in time varying frequency-selective multipath channels," \emph{Proc. IEEE Workshop Sensor Array Multichannel Signal Process.,} pp. 434-438, 2000.

\bibitem{91}
Y. Mostofi and D. C. Cox, ``ICI mitigation for pilot-aided OFDM mobile systems," \emph{IEEE Trans. Wireless Commun.,} vol. 4, no. 2, pp. 765-774, March 2005.

\bibitem{16}
H. Hijazi and L. Ros, ``Polynomial estimation of time-varying multipath gains with intercarrier interference mitigation in OFDM systems," \emph{IEEE Transactions on Vehicular Technology,} vol. 58, no. 1, pp. 140-151, Jan. 2009.

\bibitem{6}
Z. Tang, R. C. Cannizzaro, G. Leus, and P. Banelli, ``Pilot-assisted time-varying channel estimation for OFDM systems," \emph{IEEE Transactions on Signal Processing,} vol. 55, no. 5, pp. 2226-2238, May 2007.

\bibitem{42}
X. Ma, G. Giannakis, and S. Ohno, ``Optimal training for block transmissions over doubly-selective wireless fading channels," \emph{IEEE Trans. Signal Process,} vol. 51, no. 5, pp. 1351-1366,  May 2003.

\bibitem{10}
G. Taubock, F. Hlawatsch, D. Eiwen, and H. Rauhut, ``Compressive estimation of doubly selective channels in multicarrier systems: leakage effects and sparsity-enhancing processing," \emph{IEEE Journal of Selected Topics in Signal Processing,} vol. 4, no. 2, pp. 255-271, April 2010.

\bibitem{11}
G. Taubock and F. Hlawatsch, ``A compressed sensing technique for OFDM channel estimation in mobile environments: exploiting channel sparsity for reducing pilots," \emph{IEEE International Conference on Acoustics, Speech and Signal Processing (ICASSP),} pp. 2885-2888, March 2008.



\bibitem{12}
D. L. Donoho, M. Elad, and V. N. Temlyakov, ``Stable recovery of sparse overcomplete representations in the presence of noise," \emph{IEEE Trans. Inf. Theory,} vol. 52, no. 1, pp. 6-18, Jan. 2006.

\bibitem{13}
M. Elad, ``Optimized projections for compressed sensing," \emph{IEEE Transcation on Signal Processing,} vol. 55, no. 12, pp. 5695-5702, Dec. 2007.

\bibitem{14}
E. J. Candes, J. Romberg, and T. Tao, ``Robust uncertainty principles: exact signal reconstruction from highly incomplete frequency information,” \emph{IEEE Trans. Inf. Theory,} vol. 52, no. 2, pp. 489-509, Feb. 2006.

\bibitem{15}
E. J. Candes and M. B. Wakin, ``An introduction to compressive sampling," \emph{IEEE Signal Processing Mag.,} vol. 25, no. 2,  pp. 21-30, March 2008.


\bibitem{20}
X. He and R. Song, ``Pilot pattern optimization for compressed sensing based sparse channel estimation in OFDM systems," \emph{International Conference on Wireless Communications and Signal Processing (WCSP),} pp. 1-5,  Oct. 2010.

\bibitem{21}
N. Jing, W. Bi, and L. Wang, ``Deterministic pilot design for MIMO OFDM system based on compressed sensing," \emph{International Conference on Communication Technology (ICCT),} pp. 897-903, Nov. 2012.

\bibitem{22}
D. Wang and X. Hou, ``Compressed MIMO chanel estimation and efficient pilot pattern over Doppler sparse environment," \emph{International Conference on Wireless Communications and Signal Processing (WCSP),} pp. 1-5, Nov. 2011.

\bibitem{23}
C. Qi and L. Wu, ``Optimized pilot placement for sparse channel estimation in OFDM systems," \emph{IEEE Signal Processing Letters,} vol. 18, no. 12, pp. 749-752, Dec.  2011.

\bibitem{24}
C. Qi and L. Wu, ``A study of deterministic pilot allocation for sparse channel estimation in OFDM systems," \emph{IEEE Communications Letters,}  vol. 16, no. 5, pp. 742-744,  May 2012.

\bibitem{25}
X. Ren, W. Chen, and Z. Wang, ``Low coherence compressed channel estimation for high mobility MIMO OFDM  systems," \emph{Global Communications Conference (GLOBECOM),} Dec. 2013.

\bibitem{26}
X. Xiao, B. Zheng, and C. Wang, ``Compressed channel estimation based on optimized measurement matrix," \emph{Wireless Communications and Signal Processing (WCSP),} pp. 1-5, Nov. 2011.

\bibitem{9}
L. Liu, C. Tao, J. Qiu, H. Chen, L. Yu, W. Dong, and Y. Yuan, ``Position-based modeling for wireless channel on high-speed railway under a viaduct at 2.35 GHz," \emph{IEEE Journal on Selected Areas in Communications,} vol. 30, no. 4, pp. 834-845, May 2012.

\bibitem{92}
R. D. Pascoe and T. N. Eichorn, ``What is communication-based train control?," \emph{IEEE Vehicular Technology Magazine,} vol. 4, no. 4, pp. 16-21, Dec. 2009.


\bibitem{93}
H. Hijazi and L. Ros, ``Joint data QR-detection and kalman estimation for OFDM time-varying rayleigh channel complex gains," \emph{IEEE Trans. Commun.,} vol. 58, no. 1, pp. 170-178, 2010.


%
%
%
%
%
%
%








\bibitem{17}
S. S. Chen, D. L. Donoho, and M. A. Saunders, ``Atomic decompostiion by basis pursuit," \emph{SIAM Review}, vol. 43, pp. 129-159, 2001.

\bibitem{18}
Y. C. Pati, R. Rezaiifar, and P. S. Krishnaprasad, ``Orthogonal matching pursuit: recursive function approximation with applications to wavelet decomposition," \emph{Proceedings of the 27th Annual Asilomar Conference on Signals, Systems and Computers}, vol. 1, pp. 40-44, Nov. 1993.
%

\bibitem{30}
J. A. Tropp, ``Greed is good: algorithmic results for sparse approximation," \emph{IEEE Trans. Inf. Theory,}  vol. 50, no. 10, pp. 2231-2242, Oct. 2004.

\bibitem{40}
I. Berenguer, X. Wang, and V. Krishnamurthy, ``Adaptive MIMO antenna selection via discrete stochastic optimization," \emph{IEEE Trans. Signal Processing,} vol. 53, no. 11, pp. 4315-4329, Nov. 2005.

\bibitem{41}
S. Andradottir, ``A global search method for discrete stochastic optimization," \emph{SIAM J. Optimiz.}, vol. 6, no. 2, pp. 513-530,  1996.

\bibitem{43}
S. Andradottir, ``Accelerating the convergence of random search methods for discrete stochastic optimization," \emph{ACM Trans. Modeling and Compu. Simul.,} vol. 9, no. 4, pp. 349-380,  1999.

\bibitem{44}
K. Pekka, M. Juha, H. Lassi, et la., ``WINNER II Channel Models. IST-4-027756,'' \emph{WINNER II, D1.1.2 v1.1, Tech. Rep.}, Sept. 2007.

\end{thebibliography}
\end{document}